\documentclass[11pt]{article}
\usepackage[textwidth=15.2cm,textheight=22cm]{geometry}
\usepackage{amsmath,amssymb}
\usepackage{latexsym}
\usepackage{graphicx}
\usepackage{bbm}
\usepackage{bm}
\usepackage{cite}
\allowdisplaybreaks[1]
\numberwithin{equation}{section}
\newcommand{\eps}{\epsilon}

\newcommand{\be}{\begin{equation}}
\newcommand{\ee}{\end{equation}}
\newcommand{\beq}{\begin{equation}}
\newcommand{\eeq}{\end{equation}}
\newcommand\fr[1]{\frac{1}{#1}}

\def\p{\partial}

\def\bh{\bar h}
\newcommand{\nn}{\nonumber}
\newcommand{\calN}{{\mathcal N}}
\newcommand{\ndt}{\noindent}
\def\bea{\begin{eqnarray}}
\def\eea{\end{eqnarray}}
\def\beas{\begin{eqnarray*}}
\def\eeas{\end{eqnarray*}}
\def\sla{\raise.15ex\hbox{$/$}\kern-.57em}

\def\parm{{\partial}_{-}}
\def\lt{\tilde{\lambda}}
\def\spa#1.#2{\left\langle#1\,#2\right\rangle}
\def\spb#1.#2{\left[#1\,#2\right]}

\begin{document}

\begin{titlepage}
\vskip 1cm
\centerline{\Large{\bf {Gravity as the square of Yang-Mills:}}} 
\vskip 0.3cm
\centerline{\Large{\bf {implications for $\mathcal N=8$ Supergravity}}}
\vskip 1.5cm
\centerline{Sudarshan Ananth}
\vskip .5cm
\centerline{\it {Indian Institute of Science Education and Research}}
\centerline{\it {Pune 411021, India}}
\vskip 1.5cm
\centerline{\bf {Abstract}}
\vskip .5cm
\ndt The pure gravity Lagrangian can be written as the ``square" of the pure Yang-Mills Lagrangian to second order in coupling constants. This paper uses this form of the gravity Lagrangian as a starting point to arrive at a compact light-cone superspace Lagrangian for $\calN=8$ supergravity to order $\kappa^2$.
\vfill
\end{titlepage}

\section{Introduction}

Conventional quantum field theory Lagrangians are subject to field redefinitions that can drastically alter their structure and appearance. Working in light-cone gauge, where the physical degrees of freedom are manifest, combined with the use of suitable field redefinitions allows for the rewriting of these Lagrangians in forms very closely allied to on-shell physics~\cite{RM}. Although manifest locality is lost in the process, this rewriting cleans up a lot of the clutter allowing us to make the structure of scattering amplitudes manifest at the level of the Lagrangian itself. This philosophy, of staying close to on-shell physics, is used in this paper to write down a compact Lagrangian for $\calN=8$ supergravity to order $\kappa^2$. 
\vskip 0.3cm
\ndt Three of the four fundamental forces in Nature are governed by Yang-Mills Lagrangians. The fourth, gravity, is described by the very different looking Einstein-Hilbert Lagrangian. While the Yang-Mills Lagrangian terminates at finite order in the coupling constant, the gravity Lagrangian does not. Further gravity theories involve a dimensionful coupling constant that renders them non-renormalizable. It is therefore very surprising that there exist close perturbative ties between gravity theories and Yang-Mills theories. These relations, referred to as the KLT relations~\cite{KLT}, tell us that tree-level scattering amplitudes in pure gravity are the ``square" of tree-level scattering amplitudes in pure Yang-Mills. With suitable field redefinitions, these relations can be made manifest at the level of the respective Lagrangians~\cite{AT}. The light-cone Lagrangian for gravity to order $\kappa^2$ in this amplitude-friendly form reads (see appendix for notation and conventions)
\begin{eqnarray}
\label{mhvg}
&& \!\!\!\!\!\!\!{\it L}=\int d^4p\;\bh(-p)\,p^2\,h(p)+\kappa\int d^4p\;d^4k\;d^4l \nn \\
&& \!\!\!\!\!\!\!\!\!\times\,\delta^{(4)}(p\!+\!k\!+\!l)\,\frac{\spa{k}.{l}^6}{\spa{l}.{p}^2\spa{p}.{k}^2}\,h(p)\bh(k)\bh(l) \nn \\
&& \!\!\!\!\!\!\!\!\!+\kappa^2\int\! d^4p\;d^4q\;d^4k\;d^4l\,\delta^{(4)}(p\!+\!q\!+\!k\!+\!l) \nn \\
&& \!\!\!\!\!\!\!\!\!\times\,\frac{\spa{k}.{l}^8\spb{k}.{l}}{\spa{k}.{l}\spa{k}.{p}\spa{k}.{q}\spa{l}.{p}\spa{l}.{q}\spa{p}.{q}^2}\,h(p)h(q)\bh(k)\bh(l)\ .
\end{eqnarray}
where $\kappa$ is the gravitational coupling constant and $h$ and $\bh$ represent gravitons of helicity $+2$ and $-2$ respectively. We note that this form of the Lagrangian is derived from the usual light-cone Lagrangian for gravity~\cite{BCL} by a field redefinition~\cite{AT}. In this paper, we draw extensively on the result (\ref {mhvg}) to write down a suitable Lagrangian for $\calN=8$ supergravity.

\section{$\calN=8$ Supergravity}
\vskip 0.3cm
\ndt There is mounting evidence~\cite{finite} that the maximally supersymmetric $\calN=8$ supergravity is far better behaved in the ultra-violet than previously believed. If the $\calN=8$ theory is actually finite it would be the first example of an ultra-violet finite quantum field theory of gravity in four dimensions. Proving finiteness to all orders, for any theory, is no easy task. In the case of $\calN=4 $ superYang-Mills, the light-cone superspace formulation of the theory~\cite{BLN,SM} explicitly showed that all Green functions are ultra-violet finite {\it {in light-cone gauge}} (thereby implying scale-invariance, a gauge-independent statement). A light-cone superspace formulation of $\calN=8$ supergravity could prove similarly powerful and this is the primary motivation for the present paper. 
\vskip 0.3cm
\ndt The $\calN=8$ supergravity Lagrangian is known to order $\kappa^2$~\cite{BBB,ABR,ABHS} but the vertices are bulky and impossible to use, in their current form, for calculations. In this letter, we use (\ref {mhvg}) as a guide to write down a much simpler Lagrangian for the $\calN=8$ model. The main idea here is that any $\calN=8$ Lagrangian, to order $\kappa^2$, will reduce to the Lagrangian for pure gravity, to the same order (up to field redefinitions), as long as we focus on just the gravitons in the model. The justification for tracking only the gravitons in the $\calN=8$ theory stems from our use of a manifestly $\calN=8$ light-cone superfield through which the kinematical supersymmetries account for the other component fields in the model. The old gravity Lagrangian~\cite{BCL} is itself rather lengthy and hard to manipulate but the recent result (\ref {mhvg}) is extremely compact and is a good starting point. 
\vskip 0.3cm
\ndt $\calN=8$ superspace is spanned by eight Grassmann variables, $\theta^m$ and their complex conjugates ${\bar\theta}_m$ $(m = 1, ..., 8)$. The superfield describing the $\calN=8$ theory is~\cite{BLN}
\bea
&&\!\!\!\!\!\!\Phi(y)=\fr{\parm^2}h(y)+i\theta^m\fr{\parm^2}{\bar\psi}_m(y)-i\theta^{mn}\fr{\parm}{\bar A}_{mn}(y)\nn \\
&&\!\!\!\!+\theta^{mnp}\fr{\parm}\bar\chi_{mnp}(y)-\theta^{mnpq}{\bar C}_{mnpq}(y)+i{\tilde\theta}_{mnp}\chi^{mnp}(y) \nn \\
&&\!\!\!\!-i{\tilde\theta}_{mn}\parm A^{mn}(y)-{\tilde\theta}_m\parm\psi^m(y)+4\tilde\theta\parm^2\bh(y)\ ,
\eea
where 
\bea
\theta^{a_1a_2\ldots a_n}&=&\fr{n!}\,\theta^{a_1}\theta^{a_2}\cdots\theta^{a_n}\ , \nn \\
{\tilde\theta}_{a_1a_2\ldots a_n}&=&\eps_{a_1a_2\ldots a_nb_1b_2\ldots b_{(8-n)}}\,\theta^{b_1b_2\ldots b_{(8-n)}}\ ,
\eea
and $y\!=\!\!(x,{\bar x},x^+\!,y^-\!\!\equiv\!x^-\!\!-\!\frac{i}{\sqrt 2}\theta^m{\bar\theta}_m)$. We note that the component fields here differ from the usual ones~\cite{ABR} by a field redefinition, defined for the graviton in~\cite{AT}, which is the same for all the component fields (up to factors of momenta) thanks to maximal supersymmetry and our use of a $\calN=8$ superfield.

\vskip 0.3cm
\section{The $\calN=8$ Lagrangian to order $\kappa^2$}
\vskip 0.3cm
\ndt To order $\kappa^2$, the $\calN=8$ supergravity theory is described by the following Lagrangian density in light-cone superspace
\bea
\label{lag}
{\cal L}={\cal L}_0+\kappa\,{\cal L}_1+\kappa^2\,{\cal L}_2\ .
\eea
In the following we determine momentum space expressions for ${\cal L}_0$, ${\cal L}_1$ and ${\cal L}_2$. We do this by choosing an ansatz for each of these terms based on our knowledge of the $\calN=8$ theory. We then perform superspace Grassmann integrations following which, we compare the graviton portion of the resulting component Lagrangian with (\ref {mhvg}). To make such a comparison straightforward, we work entirely in momentum space. 
\vskip 0.3cm
\ndt Our ansatz for the kinetic term reads
\bea
\label{two}
{\cal L}_0=\int d^8\theta d^8\bar\theta\;\delta(p+k)\;E(p,k)\,\Phi(p)\bar\Phi(k)\ ,
\eea
where $\bar\Phi=\overline{(\Phi)}$ is the conjugate superfield and $E$ is a still to be determined function of the momenta. We focus on the gravitons in the superfield and set all remaining components of the superfield to zero. Grassmann integration followed by a comparison with (\ref {mhvg}) yields
\bea
E=\fr{32}\frac{p^2}{p_-^4}\ ,
\eea
in agreement with~\cite{BBB}. For the cubic vertex, based on the structure in~\cite{ABR}, we choose the following ansatz
\bea
\label{three}
{\cal L}_1=\int d^8\theta d^8\bar\theta\;\delta(p+k+l)\;F(p,k,l)\,\Phi(p)\bar\Phi(k)\bar\Phi(l)\ ,
\eea
with $F$ now representing the unknown momentum factor. Notice that our cubic ansatz does not contain a $\Phi\Phi\bar\Phi$ term since (\ref {mhvg}) has no $hh\bh$ term (for concerns regarding reality, we refer the reader to the discussion in~\cite{EM}). Integrating over superspace and comparing with (\ref {mhvg}), we obtain
\bea
\!\!F\!=\!-\fr{32}\frac{k_-^2l_-^2}{{(k_-\!\!+\!l_-\!)}^2{(k_-^2\!\!+\!k_-l_-\!\!+\!l_-^2)}^2}\,\times\,\frac{\spa{k}.{l}^6}{\spa{l}.{p}^2\spa{p}.{k}^2}\ ,
\eea
where momentum conservation has been used in simplifying the expression. It is important to note that these are off-shell results, valid at the level of the Lagrangian (in fact, the function $F$ vanishes on-shell). 
\vskip 0.3cm
\ndt We now turn to the quartic vertex which is trickier. Picking an ansatz here is hard because we do not have a derivation for the quartic vertex starting from the superPoincar{\' e} algebra. Here, we make an educated guess for the quartic vertex based on the existing position space Lagrangian in~\cite{ABHS}. The result in~\cite{ABHS} consists of over a hundred terms in superspace and each term, on dimensional grounds~\footnote{See section {\bf {4}} of~\cite{ABHS} for a detailed analysis.}, consists of two superfields, their conjugates and two superspace derivatives. So we start with the following ansatz for the quartic vertex
\bea
\label{quartic}
{\cal L}_2=\int d^8\theta d^8\bar\theta\;\delta(p+q+k+l)\;G(p,q,k,l) \nn \\
\times\;[{\bar q}_m\Phi(p)]\,\Phi(q)\,[q^m\bar\Phi(k)]\,\bar\Phi(l)\ ,
\eea
where $q$ and $\bar q$ refer to the supersymmetry generators~\cite{ABR} in $\calN=8$ superspace
\bea
\!\!\!\!\!\!q^m\!=\!-\frac{\p}{\p{\bar\theta}_m}-\frac{i}{\sqrt 2}\theta^m\parm\ ,\quad\!\!\! {\bar q}_m\!=\!\frac{\p}{\p\theta^m}+\frac{i}{\sqrt 2}{\bar\theta}_m\parm\ .
\eea
Their actions on the superfield and its conjugate read
\beas
{\bar q}_m\Phi(y)\!=\!+16{\sqrt 2}i{(\theta)}^8{\bar\theta}_m\parm^3h\!+\!\frac{4}{7!}\eps_{ma_2\ldots a_8}\theta^{a_2}\cdots\theta^{a_8}\parm^2h\ , \nn \\
q^m\bar\Phi(y)\!=\!-16{\sqrt 2}i{(\bar\theta)}^8\theta^m\parm^3\bh\!-\!\frac{4}{7!}\eps^{ma_2\ldots a_8}{\bar\theta}_{a_2}\cdots{\bar\theta}_{a_8}\parm^2\bh\ ,
\eeas
where we have, as before, set all components of the superfield except the graviton to zero. Using the above expressions, we Grassmann integrate (\ref {quartic}) and compare it with the last line in (\ref {mhvg}) to obtain
\bea
G\,=-\,\fr{32\sqrt 2}\frac{k_-^2l_-^2p_-q_-}{(k_-^4+l_-^4)(p_-^3+q_-^3)} \nn \\
\times\,\frac{\spa{k}.{l}^8\spb{k}.{l}}{\spa{k}.{l}\spa{k}.{p}\spa{k}.{q}\spa{l}.{p}\spa{l}.{q}\spa{p}.{q}^2}\ .
\eea
This completes our construction to order $\kappa^2$. Clearly a Lagrangian to this order does not allow us to comment on the structure (or finiteness properties) of the entire $\calN=8$ theory. However, the main message in this paper is that one should be guided exclusively by ``physical" scattering amplitudes when constructing a Lagrangian aimed at studying the ultra-violet properties of a theory.

\vskip 0.5 cm
\begin{center}
* ~ * ~ *
\end{center}
\vskip 0.1 cm

\ndt An important issue that remains to be addressed is the uniqueness of the $\calN=8$ action. Using suitable superfield redefinitions, our Lagrangian can be recast in a multitude of ways. One such superfield redefinition would relate the Lagrangian in this letter to the one in~\cite{ABHS}. This is to be expected because the two superspace expressions are based on different pure gravity Lagrangians,~\cite{ABHS} and (\ref {mhvg}). These two forms of the gravity Lagrangian are themselves related by a field redefinition (they are equal up to terms proportional to the equations of motion~\cite{AT}). Ideally, as mentioned earlier, one would like to derive this simpler $\calN=8$ superspace Lagrangian from first principles, by closing the superPoincar{\' e} algebra (like the derivation for $\calN=4$ Yang-Mills in~\cite{ABKR}). 

The light-cone Lagrangian for pure gravity was recently extended to order $\kappa^3$~\cite{SA}. In principle, this allows us to extend the result in (\ref {mhvg}) to order $\kappa^3$ using the redefinitions in~\cite{AT}. This would, in turn, determine the quintic interaction vertex in $\calN=8$ supergravity as well. We hope that this order-by-order approach will teach us something about the structure of the $\calN=8$ Lagrangian but also, more importantly, offer us a glimpse of any hidden symmetries~\cite{finite,KNR,ABKR} the theory might possess.

\subsection*{Acknowledgments}

I thank Stefano Kovacs, Hidehiko Shimada and Stefan Theisen for discussions. This work is supported by a Ramanujan Fellowship from the DST, Government of India.

\appendix 
\section{Conventions and notation}

We work with the metric $(-,+,+,+)$ and define
\be
x^\pm\,=\,\fr{\sqrt 2}\,(x^0\,\pm\,x^3)\ , \quad \partial_\pm\,=\,\fr{\sqrt 2}\,(\partial_0\,\pm\,\partial_3)\ .
\ee
$x^+$ plays the role of light-cone time and $\partial_+$ the light-cone Hamiltonian. $\parm$ is now a spatial derivative and its inverse, $\frac{1}{\parm}$, is defined using the prescription in~\cite{SM}. We define
\bea
x&&\!\!\!\!\!\!\!\!=\fr{\sqrt 2}\,(x^1\,+i\,x^2)\ , \quad {\bar \partial}\equiv\frac{\partial}{\partial x}=\fr{\sqrt 2}\,(\partial_1\,-\,i\,\partial_2)\ , \nonumber \\
{\bar x}&&\!\!\!\!\!\!\!\!=\fr{\sqrt 2}\,(x^1\,-i\,x^2)\ , \quad \partial\equiv\frac{\partial}{\partial {\bar x}}=\fr{\sqrt 2}\,(\partial_1\,+\,i\,\partial_2)\ .
\eea
\vskip -0.1cm
\ndt A four-vector $p_\mu$ may be expressed as a bispinor $p_{a \dot{a}}$ using the $\sigma^\mu=(-{\bf {1}},\,{\bf {\sigma}})$ matrices 
\beq
\label{paad}
p_{a \dot{a}}\,\equiv\,p_\mu\,{(\sigma^\mu)}_{a \dot{a}}\,=\,\left( 
 \begin{matrix}
-p_0+p_3 & \;p_1-ip_2\, \\ p_1+ip_2 & \;-p_0-p_3\, 
\end{matrix}
\right)=\sqrt{2}
\left( 
 \begin{matrix}
-p_- & \;{\overline p}\, \\ p & \;-p_+\, 
\end{matrix}
\right)
\ . 
\eeq
The determinant of this matrix is
\bea
{\mbox {det}}\,(\,p_{a \dot{a}}\,)\,=\,-2\,(\,p{\overline p}-p_+p_-\,)\,=\,-\,p^\mu p_\mu\ .
\eea
When the vector $p_\mu$ is light-like we have $p_+\,=\,\frac{p{\overline p}}{p_-}$ which is the on-shell condition. We then define spinors
\beq
\lambda_{a}\,=\,\frac{2^\fr{4}}{\sqrt p_-} 
\left( 
\begin{matrix} 
p_-  \\ 
\, -p 
\end{matrix}
\right) 
\ , 
\qquad 
\lt_{\dot{a}}\,=\,-(\lambda_a)^*\,=\,-\,\frac{2^\fr{4}}{\sqrt p_-} 
\left( 
\begin{matrix} 
p_- \\ 
\, -{\overline p} 
\end{matrix}
\right) 
\ , 
\eeq
such that $\lambda_{a}\lt_{\dot{a}}$ agrees with (\ref {paad}) on-shell. The off-shell spinor products are~\cite{QMC}
\be 
\label{product}
\spa{i}.{j}\,=\,\sqrt{2}\,\frac{p^i\,p_-^j\,-\,p^j\,p_-^i}{\sqrt{p_-^i\,p_-^j}}\ ,\qquad \spb{i}.{j}\,=\,\sqrt{2}\,\frac{{\bar p}^i\,p_-^j\,-\,{\bar p}^j\,p_-^i}{\sqrt{p_-^i\,p_-^j}}\ ,
\ee
\vskip -0.3cm
\ndt and their product is
\bea
\spa{i}.{j}\spb{j}.{i}=s_{ij}\equiv-(p_i+p_j)^2\ .
\eea

\end{document}